# A FAIR platform for reproducing mutational signature detection on tumor sequencing data

Aaron Ge[1], Tongwu Zhang[1], Clara Bodelon[1], Montserrat García-Closas[1], Jonas S. Almeida[1], Jeya B. Balasubramanian[1]

[1] National Cancer Institute Division of Cancer Epidemiology and Genetics, DCEG

**Summary**

This paper presents a portable, privacy-preserving, in-browser platform for the reproducible assessment of mutational signature detection methods from sparse sequencing data generated by targeted gene panels. The platform aims to address the reproducibility challenges in mutational signature research by adhering to the FAIR principles, making it findable, accessible, interoperable, and reusable. Our approach focuses on the detection of specific mutational signatures, such as SBS3, which have been linked to specific mutagenic processes. The platform relies on publicly available data, simulation, downsampling techniques, and machine learning algorithms to generate training data and labels and to train and evaluate models. The key achievement of our platform is its transparency, reusability, and privacy preservation, enabling researchers and clinicians to analyze mutational signatures with the guarantee that no data circulates outside the client machine.

**Availability and implementation**

Our proposed in-browser platform is publicly available under the MIT license at https://aaronge-2020.github.io/Sig3-Detection/. No data leaves this privacy-preserving environment, which can be cloned or forked and served from other domains with no restrictions. All the code and relevant data used to create this platform can be found at https://github.com/aaronge-2020/Sig3-Detection.

# 1. Introduction

Mutational signatures are patterns of somatic mutations in a tumor genome associated with specific mutagenic processes or treatments (Alexandrov et al., 2013). Detecting these signatures can provide insights into the etiology of cancer and potential therapeutic targets. For instance, mutational signature 3 (SBS3) is linked to deficiencies in the homologous recombination-based DNA damage repair pathway and can be used to identify patients eligible for PARP inhibitor treatment (Batalini et al., 2021). While whole-genome sequencing (WGS) data is required for detecting signatures accurately, targeted sequencing panel data is more readily accessible in both research and clinical settings. Therefore, improving the accuracy of mutational signature detection methods from panel data could accelerate the translational research process.

It is challenging to detect mutational signatures from targeted sequencing data, because they only cover a small fraction of the genome. Gulhan et al. (2019) proposed SigMA, a solution to detect the mutational signature associated with HR deficiency from targeted gene panels that combines various signature detection methods in a gradient-boosted decision tree. Later, Sason et al. (2021) proposed Mix-MMM, a Latent Dirichlet Allocation-based solution for signature analysis on targeted sequencing data without pre-training on rich data. While Sason et al. (2021) was unable to reproduce SigMA's performance, their report suggests that their approach outperformed it in some respects. However, in our own attempts to reproduce their performances, we were unable to obtain replicable results for either method. This motivated us to develop the platform described in this report, which aims to exemplify how researchers can improve reproducibility for future mutational signatures detection solutions. It's important to note that our models are purely discriminatory and have not been calibrated to output a probability for the patient's likelihood to have SBS3, since this is just an exercise to demonstrate replicable analysis.

Currently, developers of novel mutational signature detection methods face a reproducibility challenge due to a lack of consensus on the following items—

1) The way mutations should be classified (Koh et al. 2021).
2) The best approach for mutational signature extraction from WGS data (Omichessan et al. 2019).

3) The downsampling method used to generate panel data from WGS data.
4) The best target sequencing panels for mutational signature detection.

These problems with reproducibility reflect the need for a FAIR benchmarking framework to evaluate methods for signature detection. The FAIR guiding principles ensure that algorithms and resources used in epidemiological and clinical studies are findable, accessible, interoperable, and reusable (Garcia-Closas et al., 2022). In this report, we describe and prototype an open-source FAIR platform for benchmarking methods for mutational signature detection from panel data. We implemented our proposed FAIR platform, as shown in Figure 1A-B, to benchmark and develop methods to detect an example signature for HR deficiency, SBS3, from MSK-IMPACT panel data (Cheng et al., 2015).

**2. Materials and methods**

For this example implementation of a FAIR platform, we developed a model that detects SBS3 on breast cancer patients. First, we acquire data from the International Cancer Genome Consortium (ICGC) Data Portal using API calls. More specifically, we obtain mutational data from 623 Whole Genome Sequences (WGS) in the BRCA study of breast cancer patients, utilizing the ICGC's publicly accessible APIs.

To develop our model, we generate training data and labels. We simulate MSK-IMPACT panel sequences by downsampling the retrieved WGS data based on the regions specified by the MSK-IMPACT panel's BED file. We store these files in Google Drive for easy access. For signature labels, we perform a bootstrapped Non-Negative Least Squares (NNLS) analysis on the original WGS data, deriving relative exposure values for SBS3, which represent the percentage of mutations attributed to this particular signature.

We conduct model training and evaluation in Google Colab, a cloud-based notebook environment that fosters reproducibility. While we use Google Colab and Google Drive for convenience and accessibility, users reproducing our analysis can choose alternative storage and analysis platforms. To ensure interoperability, we use machine-readable and non-proprietary file formats such as JSON, TSV, or CSV, which are readily compatible with programming languages like Python in Google Colab.

Lastly, we deploy our models using JavaScript and WebAssembly in a web application that is version-controlled and served directly from GitHub. Our zero-footprint web application performs all computations within the client browser, eliminating data exchange with remote services. This guarantees that no data is transmitted outside the client machine, providing users with a secure, transparent, and easily reusable solution for analyzing mutational signatures. We demonstrate this for mutation signature 3. More detailed information on our methodology can be found in the supplemental materials.

## 3. Results

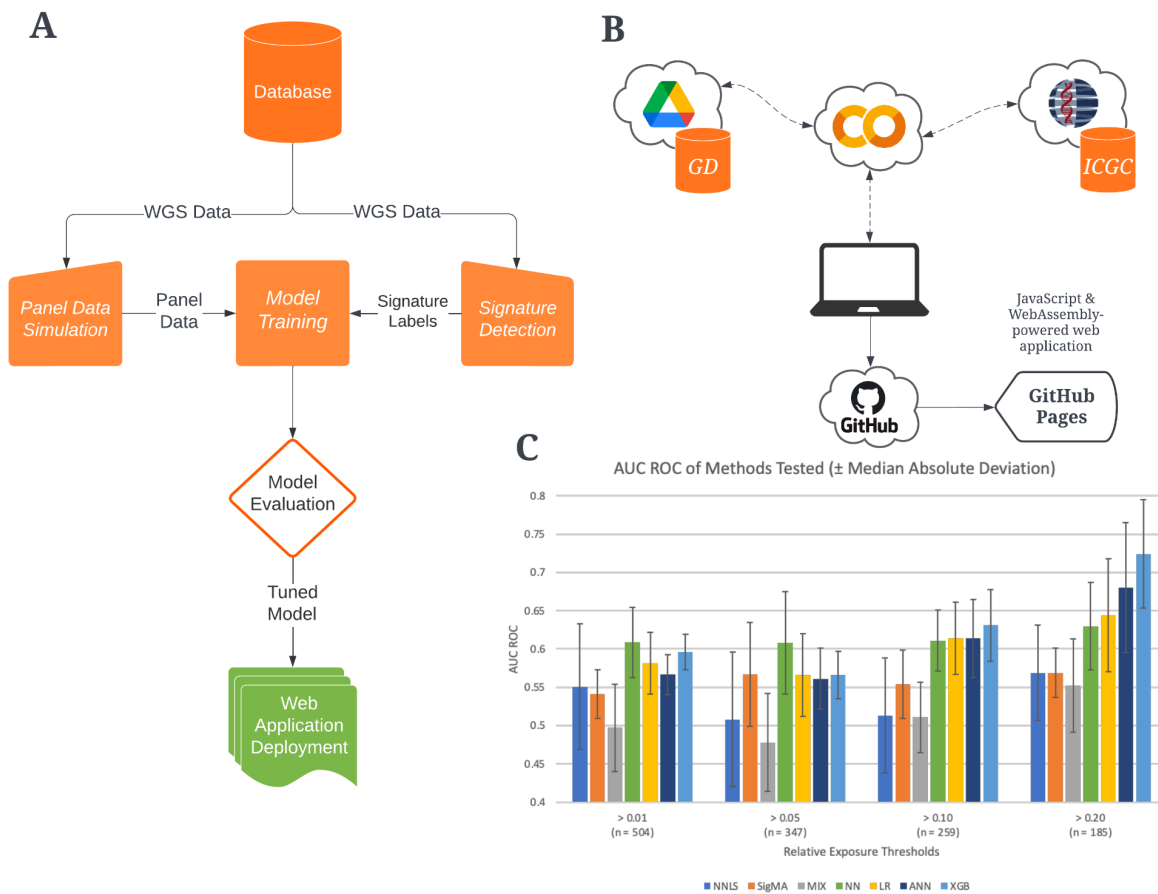

**Figure 1** depicts the workflow for developing and evaluating reproducible signature detection methods using panel data. **(A)** The workflow is hierarchical, involving obtaining whole-genome sequencing (WGS) data, training data and labels, optimizing models, and deploying them in a privacy-preserving manner. **(B)** Our FAIR (findable, reproducible, interoperable, and reusable) analysis framework involves transferring training data (not personal data, as it is part of the calibration of the models and not their use in the platform) between cloud platforms (ICGC and Google Drive), conducting the analysis in the cloud (Google Colab), and finally deploying the trained models via GitHub pages for user testing. This is the versioned hosted FAIR platform that is the focus of this report. **(C)** The results of our analysis using the FAIR workflow are shown, including the performance of various evaluated algorithms under ten-fold cross-validation, measured by the median area under the receiver operating characteristic curve (AUC ROC) at four relative exposure thresholds. We captured variance by measuring the median absolute deviation of each model's AUC ROCs under ten-fold cross-validation. The evaluated algorithms include non-negative-least-squares (NNLS), Signature Multivariate Analysis (SigMA), multinomial mixture model (MIX), nearest-neighbors (NN), logistic regression (LR), artificial neural network (ANN), and gradient-boosted decision trees (XGB). Our findings suggest that machine-learning models perform similarly to SigMA and MIX in detecting the presence of SBS3.

## 4. Conclusion

In conclusion, this report underscores the importance of reproducibility in mutational signature analysis and proposes a FAIR platform for benchmarking and developing methods for detecting mutational signatures from panel data. Our approach uses machine-readable and non-proprietary file formats, cloud-based notebook environments, and a privacy-preserving web application for model deployment. The results of our analysis suggest that machine learning models perform comparably to state-of-the-art algorithms for signature detection, demonstrating the need for easily reproducible analyses to compare and test new models.

The emergence of platforms supporting FAIR workflows, such as Signal, demonstrates the increasing importance of these principles for data-intensive analysis in epidemiology (Degasperi et al., 2020). Adherence to FAIR principles enables the development of portable, interoperable, and reusable solutions for mutational signature analysis, which could accelerate the translation of insights into clinical practice (Garcia-Closas et al., 2022).

Our proposed platform exemplifies how researchers can improve reproducibility for future mutational signature detection solutions. As the field of mutational signature analysis continues to grow, it is critical to prioritize reproducibility and ensure that methods and resources used in epidemiological and clinical studies are FAIR. By doing so, we can improve the rigor and reliability of mutational signature research and ultimately improve patient outcomes.